\documentclass[aps,pra,reprint,superscriptaddress]{revtex4-2}
\usepackage{graphicx}
\usepackage{multirow}
\usepackage{amsmath}
\usepackage[caption=false]{subfig}
\begin{document}

\title{SQUID Readout\\of a High-\(Q\) Superconducting \(LC\) Resonator}
\author{Roman Kolevatov}
\email[Corresponding author: ]{rkolevatov@princeton.edu}
\affiliation{Department of Physics, Princeton University, Princeton, New Jersey 08544, USA}

\author{Saptarshi Chaudhuri}
\affiliation{Department of Physics, Princeton University, Princeton, New Jersey 08544, USA}

\author{Lyman Page}
\affiliation{Department of Physics, Princeton University, Princeton, New Jersey 08544, USA}

\date{\today}

\begin{abstract}
We demonstrate the readout of a superconducting \(LC\) resonator, with \(Q \approx 2\times 10^{6}\) at \(\approx 250~\mathrm{kHz}\), using a dc SQUID followed by a SQUID series-array amplifier readout chain. We find that \(Q\) depends on the SQUID flux-bias point, increasing near the shallow-slope point of the SQUID modulation curve and decreasing near the steep-slope point, consistent with the previously observed SQUID damping effects. From this variation, we infer the SQUID effective input impedance. We further infer the effective temperature of the resonator circuit from the resonance peak in the noise spectrum and show that it also depends on the SQUID flux-bias point, suggesting a possible contribution from SQUID back-action noise. This system provides a working prototype for future experiments based on lumped-element resonators with SQUID readout, in particular low-mass axion searches within the DMRadio program.
\end{abstract}

\maketitle

\section{INTRODUCTION}
SQUIDs are among the most sensitive detectors of magnetic flux and are widely used in applications ranging from DC to hundreds of megahertz \cite{Clarke2004}. They are often coupled to lumped-element (\(LC\)) resonators \cite{Braginsky1985}, since the two operate in a similar frequency range, and therefore form a natural match. In such a system, the resonator enhances weak signals near resonance, while the SQUID provides highly sensitive, low-noise readout of the resulting current.

From a practical perspective, a coupled resonator--SQUID system is an excellent platform for detecting low-frequency range signals. Examples include Weber bars for gravitational wave detection \cite{Falferi2008}, low-field nuclear magnetic resonance (NMR) \cite{Sleator1987}, and resonators used in SQUID-based low-field MRI \cite{Tesche1977,Clarke1979,Clarke2007} .

In recent years, lumped-element resonators and SQUID-based readout have played an increasingly important role in low-frequency searches for low-mass axion and axionlike dark matter. Experiments employing resonant \(LC\) circuits include ADMX SLIC \cite{Crisosto2020} and the BASE Penning-trap axionlike-dark-matter search \cite{Devlin2021}, while broadband searches with a SQUID readout include ABRACADABRA-10 cm \cite{Salemi2021} and SHAFT \cite{Gramolin2021}. Most closely related to the present work are experiments that combine \(LC\) resonators with SQUID readout, such as WISPLC \cite{Zhang2022} and DMRadio \cite{SilvaFeaver2016,Brouwer2022}.

In addition to this practical aspect, the coupled SQUID--resonator is also of interest as a physical system in its own right, as it allows the study of SQUID back-action noise \cite{Hilbert1985,Martinis1986,Falferi1997,Falferi1998,Falferi2006,Falferi2009,Vinante2010,Ankel2025} and SQUID-induced damping of the resonator \cite{Hilbert1985,VanAssendelft2023}.

In this paper, we describe a superconducting lumped-element resonator with SQUID readout and a quality factor (\(Q\)) of \(Q \approx 2\times10^{6}\) at a resonance frequency of approximately 250 kHz. To our knowledge, SQUID-coupled superconducting lumped-element resonator with a comparably high \(Q\) has not previously been reported in the hundreds-of-kilohertz frequency range. For comparison, Ref.~\cite{Vinante2002} reported a resonator--SQUID system with \(Q \approx 1.1\times10^{6}\) at \(\approx 1.6\)~kHz. \(Q\) usually degrades in the hundreds-of-kilohertz range due to dielectric losses that scale approximately as the square of frequency \cite{Falferi1994}. Additional losses that increase with frequency may also arise from eddy currents in nearby normal metals. A closer comparison in frequency is the SQUID-readout resonator with \(Q \approx 4\times10^{4}\) at 492 kHz reported in Ref.~\cite{Phipps2020}.

The DMRadio search for GUT-scale QCD axions requires \(Q \ge 2\times10^{6}\) \cite{Brouwer2022}. This \(Q\) has been demonstrated without SQUID coupling \cite{Kolevatov2026}. Incorporating SQUID readout adds complexity, both because the SQUID can introduce lossy materials and because the resonator--SQUID interaction can reduce \(Q\) in ways that are difficult to predict, particularly in this previously unexplored regime. This work quantifies the interaction between the SQUID and a resonator with \(Q = 2 \times 10^6\) by determining the SQUID’s dynamic input impedance and characterizing the system’s noise performance.

The paper is organized as follows. In Sec.~\ref{sec:methods}, we describe the \(LC\) resonator and the SQUID readout. In Sec.~\ref{sec:results}, we present measurements of the flux-bias dependence of \(Q\) and resonance frequency, together with the inferred flux-bias dependence of the SQUID effective resistance and self-inductance. We also describe the effective-temperature measurements obtained from the resonance noise peak. We conclude in Sec.~\ref{sec:conclusion}. The equivalent-circuit treatment of the SQUID used to infer the SQUID effective resistance and inductance is given in Appendix~\ref{app:squid_resistance_inductance}.
\section{METHODS}
\label{sec:methods}
\begin{figure*}
    \centering
    \includegraphics[width=0.85\linewidth]{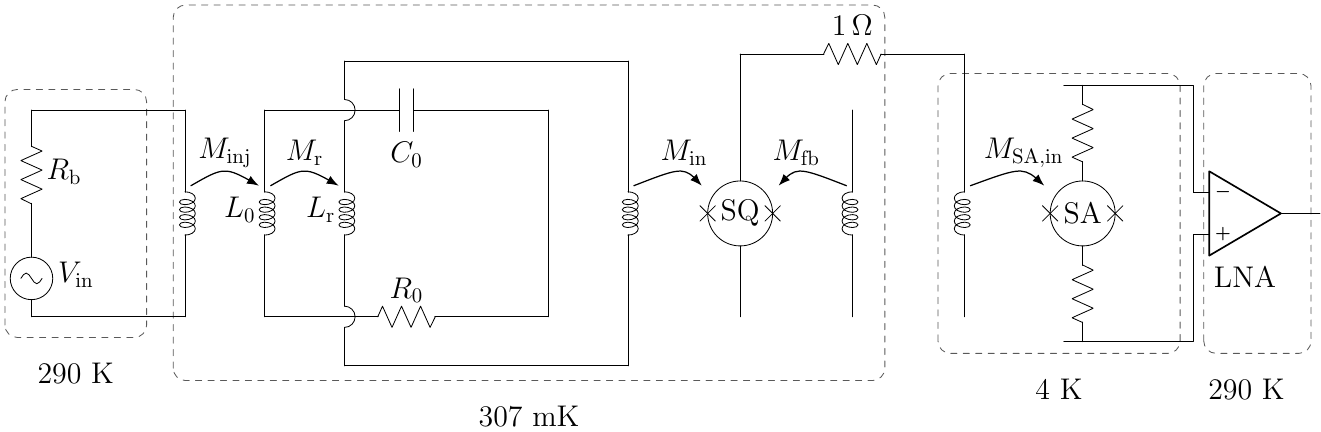}
\caption{\label{fig:circuit_diagram}
Circuit diagram of the resonator--SQUID system. The resonator consists of the inductor \(L_0\) and capacitor \(C_0\), with loss represented by the series resistance \(R_0\); the corresponding parameter values are listed in Table~\ref{tab:resonator_params}. The resonator is driven by an injection voltage \(V_{\mathrm{in}}\) applied through the bias resistor \(R_{\mathrm b}\) and coupled through the mutual inductance \(M_{\mathrm{inj}}\). The resonator is read out through a pickup loop of self-inductance \(L_{\mathrm r}\), coupled to the resonator by \(M_{\mathrm r}\) and to the SQUID input coil by \(M_{\mathrm{in}}\). The flux-bias point is tuned with a feedback coil coupled to the SQUID through \(M_{\mathrm{fb}}\). The SQUID output is connected to the \(1~\Omega\) bias resistor and to the input coil of the SQUID series-array (SA), whose output is amplified by a room-temperature LNA. Terminated lines indicate connections to components not shown. The full readout architecture, consisting of an 11-channel SQUID multiplexer followed by an SA, is described in Refs.~\cite{Doriese2016,Henderson2016}; in this work, one of the 11 SQUID channels is used. The SQUID-unbiased quality factor \(Q_0\), shown by the horizontal dashed line in Fig.~\ref{fig:Q_and_fr_dVdPhi_vs_flux}, is measured using a separate direct room-temperature readout loop that is not shown in the diagram.}
\end{figure*}
\begin{table}[t]
\caption{\label{tab:resonator_params}
Measured circuit parameters. Here \(L_0\) and \(C_0\) are the intrinsic resonator self-inductance and capacitance. The parameters \(R_0\), \(Q_0\), and \(f_{\mathrm r,0}\) are the series resistance, quality factor, and resonance frequency measured with the SQUID unbiased. The quantities \(M_{\mathrm{inj}}\), \(M_{\mathrm r}\), and \(L_{\mathrm r}\) are the injection mutual inductance, the resonator--readout mutual inductance, and the readout-loop self-inductance, respectively. The quantities \(M_{\mathrm{in}}\), \(M_{\mathrm{fb}}\), and \(M_{\mathrm{SA,in}}\) are the SQUID input-coil, feedback-coil, and SQUID series-array input-coil mutual inductances, respectively \cite{Doriese2016}. The corresponding circuit elements are shown in Fig.~\ref{fig:circuit_diagram}.}
\begin{ruledtabular}
\begin{tabular}{lc}
Parameter & Measured value \\
\hline
\(L_0\) & \((750 \pm 1)\,\mu\mathrm{H}\) \\
\(C_0\) & \((546.7 \pm 0.7)\,\mathrm{pF}\) \\
\(R_0\) & \((0.65 \pm 0.03)\,\mathrm{m}\Omega\) \\
\(Q_0\) & \((1.79 \pm 0.09)\times 10^{6}\) \\
\(f_{\mathrm r,0}\) & \((248\,552.219 \pm 0.001)\,\mathrm{Hz}\) \\
\hline
\(M_{\mathrm{inj}}\) & \((98 \pm 4)\,\mathrm{nH}\) \\
\(M_{\mathrm r}\) & \((68.85 \pm 0.02)\,\mathrm{nH}\) \\
\(L_{\mathrm r}\) & \((609.2 \pm 0.2)\,\mathrm{nH}\) \\
\hline
\(M_{\mathrm{in}}\) & \(690\,\mathrm{pH}\) \\
\(M_{\mathrm{fb}}\) & \(30\,\mathrm{pH}\) \\
\(M_{\mathrm{SA, in}}\) & \(104\,\mathrm{pH}\) \\
\end{tabular}
\end{ruledtabular}
\end{table}

\subsection{\(LC\) resonator}
A detailed description of the resonator design is given in \cite{Kolevatov2026}. Here, we provide only a brief summary before turning to the SQUID integration. The resonator apparatus is enclosed within a high-purity aluminum shield (5N), 419 mm high and 292 mm in diameter. The shield is divided into three chambers: a capacitor chamber, an inductor chamber, and a SQUID chamber. The apparatus is cooled to \(\approx 307~\mathrm{mK}\) with a \(^{3}\mathrm{He}\text{--}^{4}\mathrm{He}\) sorption refrigerator backed by pulse tubes \cite{Devlin2004,Lau2006}.

The capacitor is formed by two parallel circular aluminum 1100 plates of diameter \(197~\mathrm{mm}\), separated by a \(0.6~\mathrm{mm}\) vacuum gap. The inductor is supported by an aluminum 1100 frame, with four sapphire rods running between the two end frames. The winding consists of \(\approx\!5\) mil (\(127~\mu\mathrm{m}\)) Formvar-insulated niobium-titanium (NbTi) wire wound on the sapphire rods, forming a rectangular solenoid with cross-sectional area \(A \approx 7740~\mathrm{mm^2}\), effective length \(l \approx 121~\mathrm{mm}\), and \(N \approx 120\) turns.

The signal is injected into the resonator using a room-temperature function generator. The injection lines run from room temperature to a NbTi twisted pair inside the resonator shield. This pair forms an injection loop around one of the sapphire rods, inductively coupling the injected signal to the resonator through the mutual inductance \(M_\mathrm{inj}\).

A circuit diagram of the resonator coupled to the SQUID is shown in Fig.~\ref{fig:circuit_diagram}, and the key parameters of the resonator and readout circuit are listed in Table~\ref{tab:resonator_params}.

\subsection{SQUID readout}
To read out the signal from the resonator, a second loop of twisted NbTi pair is formed around another sapphire rod on the opposite side of the coil and coupled inductively to the resonator through the mutual inductance \(M_\mathrm{r}\). This pair is routed to the SQUID chamber, where its two ends are connected to screw terminals mounted on two aluminum 1100 blocks using tantalum screws and tantalum washers. The blocks are mounted on the SQUID printed circuit board (PCB), and aluminum wire bonds connect them to the SQUID input coil on the chip. The input coil is coupled to the SQUID with mutual inductance \(M_\mathrm{in}\).

The PCB is a custom board manufactured by Omni Circuit Boards Ltd. with superconducting aluminum traces. The SQUID chip, mounted on the PCB, is located at the center of three concentric cylindrical niobium--Cryoperm--niobium shields. Each shield is 111~mm long, and the outermost shield has an outer diameter of 23~mm. A niobium plate is attached along the full length of the underside of the SQUID PCB to provide additional electromagnetic shielding and improved thermalization. Because the PCB is lossy and also includes stainless-steel MDM connectors, we attached a vertical aluminum 1100 plate to the ceiling of the SQUID chamber. The plate divides the SQUID chamber into two sections, separating the PCB and MDM-connector side from the opposite side of the concentric shields, where the wire is routed from the inductor chamber to the SQUID chip. In this way, the plate helps reduce resonator coupling to lossy materials.

The SQUID serves as the first stage of amplification, followed by the SQUID series-array (SA) as the second stage. The SQUID--SA readout architecture follows that of Refs.~\cite{Doriese2016,Henderson2016}. The output signal is then further amplified at room temperature by a low-noise amplifier (LNA). The LNA design was provided through private communication by S. Kazuhiro of the NASA Goddard Space Flight Center. The LNA has gain \(\approx 43.2~\mathrm{dB}\), bandwidth \(\approx 31~\mathrm{MHz}\), and input-referred voltage-noise floor \(\approx 2.5~\mathrm{nV}/\sqrt{\mathrm{Hz}}\). The LNA is described in \cite{Kolevatov2024}. Depending on the particular measurement, the LNA output is further amplified using additional commercial amplifiers from Stanford Research Systems (SR560, SR445, or both).

\begin{figure}[t]
    \centering

    \subfloat[\label{fig:mod_curve}]{%
        \includegraphics[width=0.82\columnwidth]{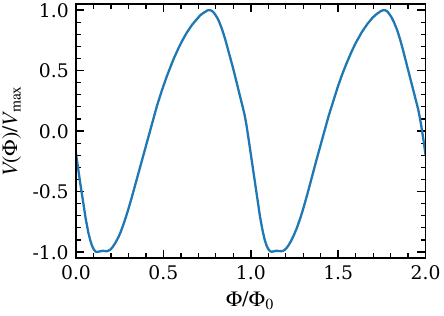}%
    }

    \vfill

    \subfloat[\label{fig:squid_noise}]{%
        \includegraphics[width=0.82\columnwidth]{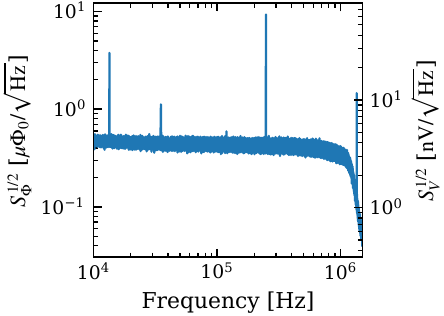}%
    }

\caption{\label{fig:mod_and_noise}
(a) Smoothed SQUID modulation curve, \(V(\Phi)\), normalized to its maximum value \(V_{\max}\). The horizontal axis shows the flux applied to the SQUID through the feedback coil. As expected, the response is periodic in flux, with a period of one magnetic flux quantum, \(\Phi_0\). The vertical axis is normalized to span from \(-1\) to \(1\), and the curve is shown over two flux-ramp periods.
(b) Noise amplitude spectrum of the SQUID+SA, referred to the SQUID input at the shallow-slope flux-bias point of \(V(\Phi)\) in flux units (left axis) and to the LNA input in voltage units (right axis). At this bias point, \(\left|dV/d\Phi\right|_{\mathrm{norm}}\approx 0.6\) is locally maximal. The spectrum spans \(10~\mathrm{kHz}\) to \(1.5~\mathrm{MHz}\) and was obtained by averaging 100 traces, each recorded at \(f_{\mathrm{s}}=3~\mathrm{MHz}\) with \(3\times1024\times1024\) samples. The data are shown on log--log axes. The roll-off near \(1.5~\mathrm{MHz}\) is set by the anti-aliasing filter.}
\end{figure}

The SQUID modulation curve, \(V(\Phi)\), gives the SQUID voltage response as a function of input magnetic flux. The measured \(V(\Phi)\) is shown in Fig.~\ref{fig:mod_curve}. As expected, the modulation is periodic in flux with period \(\Phi_0\), where \(\Phi_0 = h/(2e) \approx 2.07\times10^{-15}\,\mathrm{Wb}\) is the magnetic flux quantum. The SQUID operating points are chosen at flux values where the SQUID response magnitude \(\left|dV/d\Phi\right|\) is largest. These points are set by the DC flux applied through the feedback coil, which couples to the SQUID through the mutual inductance \(M_\mathrm{fb}\) shown in Fig.~\ref{fig:circuit_diagram}. Throughout this work, we refer to these points as \textit{flux-bias points}. Each period of \(V(\Phi)\) contains two such points, corresponding to the steep negative and shallow positive slopes of \(V(\Phi)\), where \(\left|dV/d\Phi\right|\) reaches local maxima. When the SQUID is biased at one of these points, an input flux \(\Phi_\mathrm{in}\), coupled to the SQUID through the mutual inductance \(M_\mathrm{in}\) shown in Fig.~\ref{fig:circuit_diagram}, is converted to an output voltage with gain \(\left|dV/d\Phi\right|\), provided that the SQUID remains in the linear regime, i.e., \(\Phi_\mathrm{in} \ll \Phi_0\). In general, the measured \(\left|dV/d\Phi\right|\) also includes the gain of the amplification stages following the SQUID. To avoid ambiguity, \(\left|dV/d\Phi\right|\) is normalized to unity throughout this work by dividing it by its value at the steep-slope flux-bias point, where \(\left|dV/d\Phi\right|\) is maximal.

The noise spectrum referred to the LNA input, measured with the SQUID biased at the shallow-slope flux-bias point of \(V(\Phi)\), is shown in Fig.~\ref{fig:squid_noise}. Throughout this work, the measured noise is referred to the SQUID input flux by dividing the voltage noise by non-normalized \(\left|dV/d\Phi\right|\) at the relevant flux-bias point. Within the displayed frequency range, only a few spurious peaks are visible, which correspond to electromagnetic interference (EMI). The peak near \(250~\mathrm{kHz}\) is the resonance. This indicates that the resonance is read out through the SQUID--SA chain without significant noise contamination.

\section{RESULTS}
\label{sec:results}
\subsection{\(Q\) dependence on flux-bias point}
\label{sec:Q_dep_fb_point}
\begin{figure}
    \centering
    \includegraphics[width=1\columnwidth]{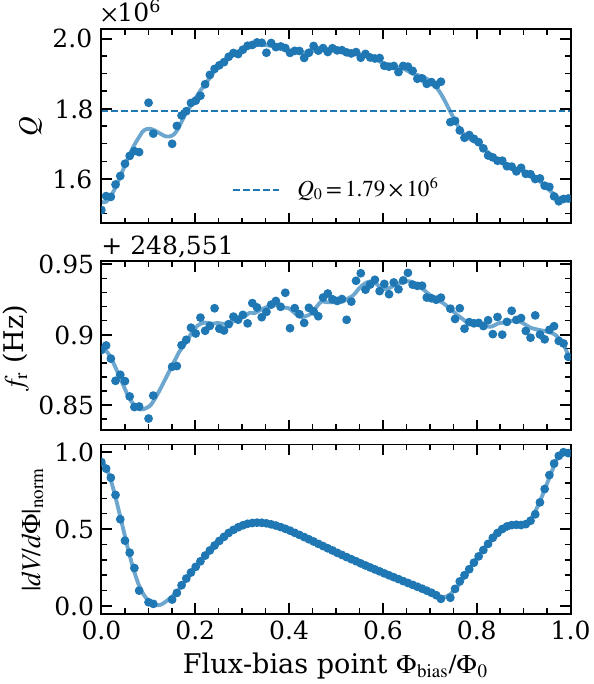}
\caption{
Resonator quality factor \(Q\) (top), resonance frequency \(f_{\mathrm r}\) (middle), and normalized SQUID response magnitude \(\left|dV/d\Phi\right|_{\mathrm{norm}}\) (bottom) versus feedback flux at \(T=307~\mathrm{mK}\), with a calibration-curve fit uncertainty of approximately 1 mK. The data are shown over one SQUID modulation period, from \(0\) to \(\Phi_0\); the periodicity was verified by extending the measurement to \(2\Phi_0\), not shown here. The dashed horizontal line marks the SQUID-unbiased value \(Q_0=1.79\times 10^6\). The semi-transparent thick curves show Savitzky--Golay-smoothed trends obtained from the measured data.
}
\label{fig:Q_and_fr_dVdPhi_vs_flux}
\end{figure}
\begin{figure}
    \centering
    \includegraphics[width=1\columnwidth]{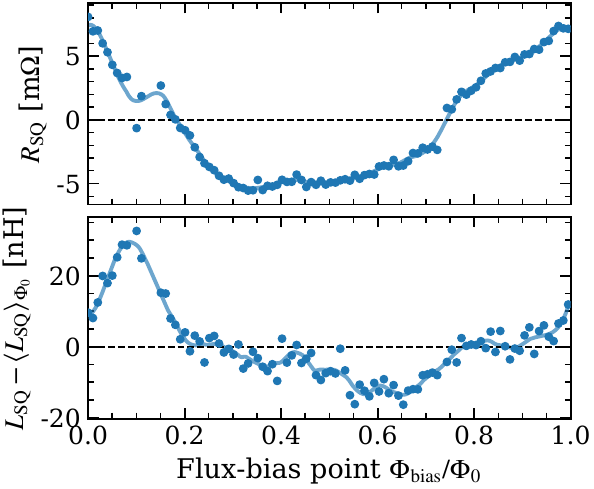}
\caption{
SQUID resistance \(R_{\mathrm{SQ}}\) (top), and SQUID self-inductance with the average over one flux period subtracted,
\(L_{\mathrm{SQ}}-\langle L_{\mathrm{SQ}}\rangle_{\Phi_0}\) (bottom) versus feedback flux at \(T=307~\mathrm{mK}\), shown over one SQUID modulation period from \(0\) to \(\Phi_0\). The periodicity was verified by extending the measurement to \(2\Phi_0\), not shown here. The semi-transparent thick curves show Savitzky--Golay-smoothed trends obtained from the measured data.
}
\label{fig:Lsq_and_Rsq_vs_flux}
\end{figure}

The dependence of \(Q\) on the SQUID flux-bias point was measured by sweeping the DC current in the feedback coil from \(0\) to approximately \(140~\mu\mathrm{A}\), corresponding to an applied SQUID feedback flux from \(0\) to about \(2\,\Phi_0\), sampled at 200 equally spaced points. At each point, \(\left|dV/d\Phi\right|\) was measured using a small sinusoidal calibration signal applied to the feedback coil, corresponding to approximately \(0.02\,\Phi_{0\mathrm{pp}}\), so as to keep the SQUID response in the linear regime. Ringdown traces were then acquired to extract the average \(Q\).

For the ringdown measurement, a current of approximately \(1~\mu\mathrm{A}\) was injected into the resonator injection circuit (left in Fig.~\ref{fig:circuit_diagram}) at \(248\,510~\mathrm{Hz}\), about \(42~\mathrm{Hz}\) de-tuned from resonance. The readout signal was amplified, mixed down to \(1~\mathrm{kHz}\) using a Mini-Circuits ZAD-6+ mixer \cite{MiniCircuits_ZAD6}, and filtered with 1-kHz high-pass and low-pass stages before extracting the ringdown relaxation time \(\tau\). The mixer was operated well below its 1-dB compression point. The room-temperature gain settings and the number of ringdown traces were adjusted adaptively according to the measured \(\left|dV/d\Phi\right|\). \(Q\) was obtained from \(\tau\) following Ref.~\cite{Kolevatov2026}. Points with normalized \(\left|dV/d\Phi\right|_\mathrm{norm}\lesssim 0.01\) were excluded because of poor signal-to-noise ratio. The SQUID feedback flux and burst injection were both generated using Agilent EDU33212A function generators. The results are shown in Fig.~\ref{fig:Q_and_fr_dVdPhi_vs_flux}.

After completion of the sweep, we also measured \(Q\) with the SQUID unbiased. From five coherently averaged ringdown traces, we obtained \(Q \approx 1.8\times10^6\), shown by the dashed line in the top panel of Fig.~\ref{fig:Q_and_fr_dVdPhi_vs_flux}.

The data show a clear correlation between \(Q\) and the SQUID flux-bias point, as shown in the top panel of Fig.~\ref{fig:Q_and_fr_dVdPhi_vs_flux}.\(Q\) varies periodically with flux with period \(\Phi_0\) and remains within the range \((1.5\text{--}2.0)\times 10^6\). This range is centered near the value measured with the SQUID unbiased, \(Q_0 \approx 1.8\times 10^6\). The steep-slope flux-bias point of the SQUID modulation curve \(V(\Phi)\), corresponding to the larger local maximum of \(\left|dV/d\Phi\right|_{\mathrm{norm}}\), is associated with a lower \(Q\), consistent with the SQUID-induced damping reported in Ref.~\cite{VanAssendelft2023}. In contrast, the shallow-slope flux-bias point, corresponding to the smaller local maximum of \(\left|dV/d\Phi\right|_{\mathrm{norm}}\), is associated with a higher \(Q\), suggesting that coupling to the SQUID can effectively reduce the net dissipation of the resonator.

The measured resonance frequency also shows a clear dependence on flux bias, as shown in the middle panel of Fig.~\ref{fig:Q_and_fr_dVdPhi_vs_flux}. The overall peak-to-peak variation of the resonance frequency is approximately \(111~\mathrm{mHz}\), with the minimum aligned with the minimum of \(\left|dV/d\Phi\right|_{\mathrm{norm}}\).

Note that the fractional change in \(Q\), \(\delta Q/Q_0\simeq 0.3\), is much larger than the fractional frequency shift, \(\delta f_{\mathrm r}/f_{\mathrm r,0}\simeq 2\times 10^{-7}\). At each flux-bias point, we define the measured variations in \(Q\) and \(f_{\mathrm r}\) as
\begin{align*}
\delta Q &= Q - Q_0,\\
\delta f_{\mathrm r} &= f_{\mathrm r} - f_{\mathrm r,0},
\end{align*}
where \(Q_0\) and \(f_{\mathrm r,0}\) are, respectively, the quality factor and resonance frequency measured with the SQUID unbiased and listed in Table~\ref{tab:resonator_params}. These variations are used to infer the SQUID resistance and self-inductance, \(R_{\mathrm{SQ}}\) and \(L_{\mathrm{SQ}}\), using the effective-circuit treatment described in Appendix~\ref{app:squid_resistance_inductance}, following the approach of Ref.~\cite{VanAssendelft2023}. In the small-perturbation limit, these quantities are given by
\begin{subequations}
\nonumber
\begin{align}
R_{\mathrm{SQ}}
&\simeq
-\frac{L_{\mathrm r}^2}{M_{\mathrm r}^2}
R_0
\frac{\delta Q}{Q_0},\\
L_{\mathrm{SQ}}
&\simeq
-2\frac{L_{\mathrm r}^2}{M_{\mathrm r}^2}
\left(
L_0-\frac{M_{\mathrm r}^2}{L_{\mathrm r}}
\right)
\frac{\delta f_{\mathrm r}}{f_{\mathrm r,0}},
\end{align}
\end{subequations}
where \(L_0\), \(R_0\), \(Q_0\), \(f_{\mathrm r,0}\), \(L_\mathrm{r}\) and \(M_\mathrm{r}\) are given in Table~\ref{tab:resonator_params}. The inferred SQUID resistance and inductance are shown in Fig.~\ref{fig:Lsq_and_Rsq_vs_flux}.

The SQUID resistance, shown in the top panel of Fig.~\ref{fig:Lsq_and_Rsq_vs_flux}, varies from approximately \(-6\) to \(8~\mathrm{m}\Omega\) over one \(\Phi_0\) period. It is positive near the steep-slope flux-bias point of the SQUID modulation curve \(V(\Phi)\), corresponding to the larger local maximum of \(\left|dV/d\Phi\right|_{\mathrm{norm}}\), crosses zero near the point where \(\left|dV/d\Phi\right|_{\mathrm{norm}}=0\), and becomes negative near the shallow-slope flux-bias point, corresponding to the smaller local maximum of \(\left|dV/d\Phi\right|_{\mathrm{norm}}\). This sign change is consistent with the observed modulation of \(Q\): positive \(R_{\mathrm{SQ}}\) increases the effective dissipation and lowers \(Q\), whereas negative \(R_{\mathrm{SQ}}\) reduces the net dissipation and increases \(Q\). Near \(R_{\mathrm{SQ}}=0\), \(Q\) is close to the value measured with the SQUID unbiased.

The SQUID self-inductance is shown in the bottom panel of Fig.~\ref{fig:Lsq_and_Rsq_vs_flux}. Its peak-to-peak variation is approximately \(44~\mathrm{nH}\) over one \(\Phi_0\) period, with the maximum occurring near the minimum of \(\left|dV/d\Phi\right|_{\mathrm{norm}}\). The inferred inductance oscillates about its average over one flux period,
\begin{equation}
\left\langle L_{\mathrm{SQ}}\right\rangle_{\Phi_0}
=
\frac{1}{\Phi_0}
\int_{0}^{\Phi_0}
L_{\mathrm{SQ}}(\Phi)\,d\Phi .
\end{equation}
We estimate this average to lie in the range
\begin{equation}
-33~\mathrm{nH}
\lesssim
\left\langle L_{\mathrm{SQ}}\right\rangle_{\Phi_0}
\lesssim
16~\mathrm{nH},
\end{equation}
as discussed in Appendix~\ref{app:squid_resistance_inductance} and summarized in Eq.~\eqref{eq:Lsq_boundaries}. This estimate follows from the requirement that \(L_{\mathrm{SQ}}\) cross zero within one flux period \cite{Hilbert1985}.

\subsection{Effective temperature from the resonance noise peak}
\begin{figure}
    \centering
    \includegraphics[width=0.90\columnwidth]{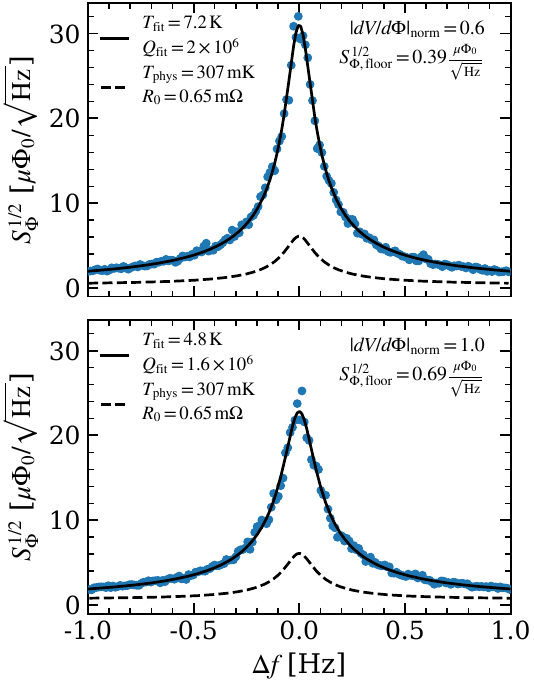}
    \caption{
Flux-noise amplitude spectrum of the SQUID-noise resonance peak, referred to the SQUID input and measured at two local maxima of \(\left|dV/d\Phi\right|\). The top and bottom panels correspond to the shallow-slope, \(\left|dV/d\Phi\right|_{\mathrm{norm}}\approx 0.6\), and steep-slope, \(\left|dV/d\Phi\right|_{\mathrm{norm}}=1\), flux-bias points of the SQUID modulation curve \(V(\Phi)\), respectively, shown in Fig.~\ref{fig:mod_curve}. The physical temperature of the copper plate on which the resonator was mounted, determined from calibrated ruthenium oxide (ROX) thermometer readings, was \(T=307~\mathrm{mK}\), with an approximately \(1~\mathrm{mK}\) calibration-curve fit uncertainty. In each panel, the blue points show the measured spectrum, calculated as the square root of the mean power spectral density (PSD) of the retained traces. Traces were retained after a \(2\sigma\) median absolute deviation (MAD) outlier rejection was applied independently to the fitted values of \(Q\), \(f_{\mathrm r}\), \(T_{\mathrm{eff}}\), and the white-noise floor \(S_{\Phi,\mathrm{floor}}^{1/2}\). Before averaging, each trace was shifted to the frequency offset \(\Delta f = f-f_{\mathrm r}\), where \(f_{\mathrm r}\) is the fitted resonance frequency of that trace. The solid black curve shows the fit to the model in Eq.~\eqref{eq:spectrum_model}, with a fixed white-noise floor \(S_{\Phi,\mathrm{floor}}^{1/2}\) added in quadrature. The dashed black curve shows the expected response for \(T_{\mathrm{phys}}=307~\mathrm{mK}\) and the SQUID-unbiased resonator quality factor \(Q_0\), equivalent to \(R_0 \approx 0.65~\mathrm{m}\Omega\). The extracted \(Q\) values differ between the two flux-bias points, consistent with Fig.~\ref{fig:Q_and_fr_dVdPhi_vs_flux}, while the fitted effective temperatures are approximately \(7~\mathrm{K}\) and \(5~\mathrm{K}\) for the shallow- and steep-slope points, respectively.
}
\label{fig:resonance_peak_mean_asd_fit}
\end{figure}
In a separate measurement, we observed the resonance noise peak of the resonator with the SQUID--SA readout chain at the two flux-bias points on the SQUID modulation curve corresponding to the local maxima of the response magnitude \(\left|dV/d\Phi\right|\).

Since this peak follows the spectrum of Johnson--Nyquist current noise in the resonator circuit, its amplitude provides an independent estimate of the effective temperature \(T_\mathrm{eff}\) of the resonator. The main challenge, however, is the narrow linewidth of the high-\(Q\) resonance, \(\Delta f = f_{\mathrm r,0}/Q_0 \approx 139~\mathrm{mHz}\), which requires a frequency resolution well below \(\Delta f\) and therefore a relatively long acquisition time for a single trace. The measurement is further complicated by the high sensitivity of the SQUID to external EMI.

Special care was taken to mitigate noise. The SQUID and SA biasing, as well as the setting of the flux-bias point, were implemented using analog circuits with potentiometers and NiMH batteries. The entire cryostat was enclosed in a copper-mesh Faraday cage, with an aluminum 6061 plate placed underneath. All unnecessary electronics were turned off.

The SQUID response magnitude \(\left|dV/d\Phi\right|\) was first measured over one full period of the SQUID modulation curve, in order to identify the two flux-bias points corresponding to its local maxima. The SQUID was then set to the desired flux-bias point, corresponding to one of the local maxima of \(\left|dV/d\Phi\right|\). To improve frequency resolution and signal-to-noise ratio, resonance-noise traces were acquired continuously over \(2.2~\mathrm{h}\) at a sampling rate of \(3~\mathrm{MHz}\), with \(N = 256\times1024\times1024\) samples per trace, corresponding to a frequency resolution of approximately \(11~\mathrm{mHz}\) and an acquisition time of about \(89~\mathrm{s}\) per trace. For each trace, only the spectral window within \(\pm 1.25~\mathrm{kHz}\) of the resonance was retained. Each trace was then converted to SQUID-input flux units using the measured value of \(\left|dV/d\Phi\right|\) and fitted to the model spectrum:
\begin{equation}
\label{eq:spectrum_model}
S_{\Phi,\mathrm{model}}^{{1/2}}(f)
=
\sqrt{\frac{4 k_B T_\mathrm{eff}}{R}}\,
\frac{M_{\mathrm{in}} M_r/{L_{r}}}{\sqrt{1 + \left(2Q\frac{f-f_{\mathrm r}}{f_{\mathrm r}}\right)^2}},
\end{equation}
where \(R = 2\pi f_{\mathrm r} L/Q\) is the effective series resistance of the resonator; see Eq.~\eqref{eq:Reff}. Here \(Q\), \(f_{\mathrm r}\), and \(T_\mathrm{eff}\) are fitting parameters corresponding to the quality factor, resonance frequency, and effective temperature of the resonator, respectively; \(k_B\) is the Boltzmann constant; and the circuit parameter values are given in Table~\ref{tab:resonator_params}. The noise floor was added in quadrature to Eq.~\eqref{eq:spectrum_model} and was estimated from the median white-noise level of the spectrum outside a \(\pm 500~\mathrm{Hz}\) window around the resonance. This procedure yielded fitted values of \(Q\), \(f_{\mathrm r}\), \(T_{\mathrm{eff}}\), and the noise floor for each trace. The resulting data set was filtered by rejecting outlier traces using a \(2\sigma\) MAD criterion applied independently to the fitted values of \(Q\), \(f_{\mathrm r}\), \(T_{\mathrm{eff}}\), and the noise floor. The retained traces were then aligned by their fitted resonance frequencies and averaged in PSD, after which the corresponding mean amplitude spectrum was obtained. This mean spectrum was fitted again using Eq.~\eqref{eq:spectrum_model}, with the noise floor added in quadrature, to extract the final values of \(Q\), \(f_{\mathrm r}\), and \(T_{\mathrm{eff}}\). The resulting mean spectrum and fitted model are shown in Fig.~\ref{fig:resonance_peak_mean_asd_fit}. The measured \(Q\) values agree well with those obtained from the ringdown measurements at the corresponding flux-bias points, as shown in Fig.~\ref{fig:Q_and_fr_dVdPhi_vs_flux}.

The inferred effective temperature is significantly higher than the resonator physical temperature of approximately \(307~\mathrm{mK}\), measured by ROX mounted on top of the resonator shield. There are a number of possible explanations for the high effective temperature. First, the effective temperature depends on the flux-bias point, suggesting a possible contribution from SQUID back-action noise \cite{Tesche1979}, independent of the excess noise from \(R_{\mathrm{SQ}}\) \cite{Hilbert1985}. 

Second, to reduce EMI pickup, we added a cold current filter at the resonator input. The filter acts as a passive current divider, reducing the injected current, and therefore pickup current noise by a factor of approximately 200. With the filter installed, \(Q\) was reduced to \(Q\simeq (9.7\pm0.3)\times 10^5\), and the flux-bias dependence of both \(Q\) and \(T_{\mathrm{eff}}\) disappeared. The effective temperature was reduced to \(T_{\mathrm{eff}}\simeq 2~\mathrm{K}\), suggesting that EMI noise was suppressed.

Third, there may be some microphonic coupling to the \(LC\) resonator and readout, which manifests as noise.

Fourth, some of the remaining excess temperature may indicate incomplete thermalization of the resonator, possibly associated with the superconducting NbTi wire (\(T_{\mathrm c}\approx 9.2~\mathrm{K}\)), since thermal transport below \(T_{\mathrm c}\) is phonon dominated \cite{Bardeen1959}, with the electronic contribution becoming negligible below about \(0.3T_{\mathrm c}\approx 2.7~\mathrm{K}\) \cite{Flachbart1978}.


\section{CONCLUSION}
\label{sec:conclusion}
In this paper, we have demonstrated the integration of a SQUID--SA readout chain with a high-\(Q\) superconducting resonator operating in the \(Q \approx 2\times10^6\) regime. This integration constitutes the central result of the work. We show that, even after coupling the resonator to the SQUID, which introduces additional lossy materials and can exert damping on the resonator, \(Q\) remains sufficiently high to enable subsequent integration of a similar system into a low-mass axion search experiment \cite{Brouwer2022}, in particular DMRadio-50L \cite{Rapidis2022,DMRadio50LDesign}.

We observe a strong dependence of \(Q\) and \(f_{\mathrm r}\) on the flux-bias point, as shown in Fig.~\ref{fig:Q_and_fr_dVdPhi_vs_flux}. In particular, \(Q\) reaches approximately \(2.0\times10^6\) at the shallow-slope flux-bias point and is reduced to approximately \(1.5\times10^6\) at the steep-slope flux-bias point; this behavior is periodic, with period \(\Phi_0\). Importantly, the resonator maintains a high \(Q\), \((1.5\text{--}2.0)\times10^6\), over the full range of SQUID flux-bias points over one period of the SQUID modulation curve.

From the observed variations \(\delta Q\) and \(\delta f_{\mathrm r}\) at each flux-bias point, we inferred the SQUID resistance \(R_{\mathrm{SQ}}\) and self-inductance \(L_{\mathrm{SQ}}\), shown in Fig.~\ref{fig:Lsq_and_Rsq_vs_flux}.

From the measured resonance noise peak of the resonator--SQUID--SA system, several main observations emerge. First, the effective temperature depends on the SQUID flux-bias point, suggesting a contribution from SQUID back-action noise. Second, after adding a current filter, the inferred effective temperature is reduced, indicating that part of the excess temperature can be attributed to EMI. Third, microphonic coupling to the \(LC\) resonator and readout may also contribute excess noise. Finally, part of the elevated temperature could be due to imperfect thermalization of the superconducting NbTi wire, owing to its poor thermal conductivity below \(T_{\mathrm c}\).

Future work will focus on replacing aluminum 1100 with niobium in the inductor coil frame and capacitor plates, because niobium provides a better thermal-contraction match to the NbTi wire and may improve thermalization. Future work will also address the remaining excess noise. Finally, this work will inform the near-term design and implementation of the SQUID readout for the DMRadio-50L resonator \cite{Rapidis2022} and, over the longer term, the development of SQUID readout systems for CAL-Pathfinder, DMRadio-Core \cite{Ankel2026}, and DMRadio-GUT \cite{Brouwer2022,Kuenstner2025}.

\begin{acknowledgments}
We thank Kent Irwin, Lindley Winslow, Chiara Salemi, Maria Simanovskaia, Aya Keller, and Victoria Ankel for their careful reading of the manuscript and valuable comments. We thank Bert Harrop for help with the SQUID chip and for wire-bonding the SQUID PCB, and Stanley Chidzik for assistance with the room-temperature low-noise amplifier. We thank Glenn Atkinson and Matt Komor for machining parts of the SQUID chamber. We also thank the undergraduates Deniz Erdag and Nicky He for assistance in designing the SQUID PCB. We thank members of the DMRadio and PXS collaborations for useful discussions. This material was based upon work supported by the U.S. Department of Energy, Office of Science, Office of High Energy Physics, under Award Number DE-SC0007968. This work has been supported by Simons Foundation grant number MP-TMPS-00002996 and the ``Table-top experiments for fundamental physics'' program, sponsored by the Gordon and Betty Moore Foundation, Simons Foundation, Alfred P. Sloan Foundation, and John Templeton Foundation. We gratefully acknowledge support from the Princeton Innovation Fund for New Ideas in the Natural Sciences. S.C. was supported by the R.H. Dicke Postdoctoral Fellowship.
\end{acknowledgments}
\appendix
\section{SQUID Resistance and Self-Inductance}
\label{app:squid_resistance_inductance}
\begin{figure}[t]
    \centering

    \subfloat[\label{fig:squid_equivalent_circuit}]{%
        \includegraphics[width=0.75\columnwidth]{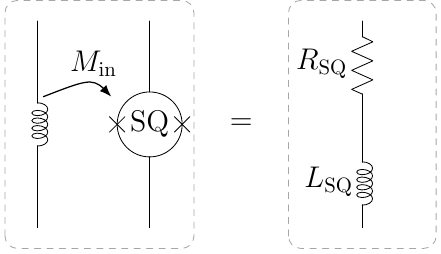}%
    }

    \vfill

    \subfloat[\label{fig:resonator_squid_equivalent_circuit}]{%
        \includegraphics[width=0.75\columnwidth]{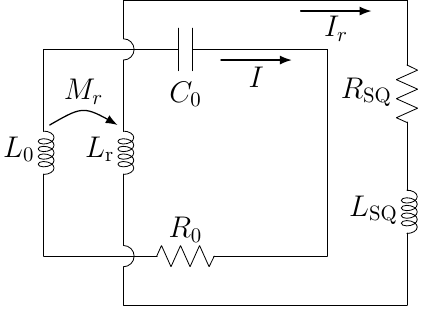}%
    }

\caption{\label{fig:appendix_equivalent_circuits}
(a) Equivalent-circuit representation of the SQUID in the readout loop as a flux-dependent resistance \(R_{\mathrm{SQ}}\) and self-inductance \(L_{\mathrm{SQ}}\).
(b) Circuit diagram of the resonator; cf. Fig.~\ref{fig:circuit_diagram}. The resonator consists of an inductor \(L_0\), capacitor \(C_0\), and series resistance \(R_0\), and is coupled through the mutual inductance \(M_{\mathrm{r}}\) to a pickup loop with self-inductance \(L_{\mathrm{r}}\). The SQUID in the pickup loop is represented by the flux-dependent resistance \(R_{\mathrm{SQ}}\) and self-inductance \(L_{\mathrm{SQ}}\).}
\end{figure}

In this Appendix, we describe an equivalent-circuit treatment of the SQUID in the readout loop, represented by flux-dependent resistance and self-inductance terms \(R_{\mathrm{SQ}} = R_{\mathrm{SQ}}(\Phi)\) and \(L_{\mathrm{SQ}}=L_{\mathrm{SQ}}(\Phi)\) coupled to the resonator. A similar effective-circuit approach was proposed in Ref.~\cite{VanAssendelft2023}; however, that work considered a different input-circuit configuration. It allows us to infer \(R_{\mathrm{SQ}}\) and \(L_{\mathrm{SQ}}\) from the measured \(Q\) and \(f_{\mathrm r}\), respectively, using the ringdown measurements discussed in Sec.~\ref{sec:Q_dep_fb_point}.

With this representation, the resonator is inductively coupled to a readout loop containing the flux-dependent resistance \(R_{\mathrm{SQ}}\) and self-inductance \(L_{\mathrm{SQ}}\), as shown in Fig.~\ref{fig:resonator_squid_equivalent_circuit}. The readout-loop current is
\begin{equation*}
I_{\mathrm r}
=
-\frac{i\omega M_{\mathrm r} I}
{R_{\mathrm{SQ}}+i\omega \left(L_{\mathrm r}+L_{\mathrm{SQ}}\right)} .
\end{equation*}
Substituting this expression into the resonator-loop equation gives the resonator effective impedance
\begin{equation*}
Z = R_0 + \frac{1}{i\omega C_0} + i\omega L_0 + \frac{\omega^2 M_{\mathrm r}^2}{R_{\mathrm{SQ}}+i\omega \left(L_{\mathrm r}+L_{\mathrm{SQ}}\right)} .
\end{equation*}
Separating the last term into real and imaginary parts and taking the limit \(R_{\mathrm{SQ}}\ll \omega\left(L_{\mathrm r}+L_{\mathrm{SQ}}\right)\), which is verified a posteriori below, we obtain
\begin{equation}
\label{eq:Z}
\begin{aligned}
Z \simeq{}& R_0 + \frac{M_{\mathrm r}^2}{\left(L_{\mathrm r}+L_{\mathrm{SQ}}\right)^2}R_{\mathrm{SQ}} + \frac{1}{i\omega C_0}\\
&+ i\omega L_0 - i\omega \frac{M_{\mathrm r}^2}{L_{\mathrm r}+L_{\mathrm{SQ}}}.
\end{aligned}
\end{equation}
Thus, the coupled readout loop contributes an effective series resistance and shifts the resonator inductance according to
\begin{subequations}
\label{eq:Reff_Leff}
\begin{align}
R &= R_0 + \frac{M_{\mathrm r}^2}{\left(L_{\mathrm r}+L_{\mathrm{SQ}}\right)^2}R_{\mathrm{SQ}},
\label{eq:Reff}
\\*
L &= L_0 - \frac{M_{\mathrm r}^2}{L_{\mathrm r}+L_{\mathrm{SQ}}}.
\end{align}
\end{subequations}
Because \(R_{\mathrm{SQ}}\) and \(L_{\mathrm{SQ}}\) are flux dependent, Eqs.~\eqref{eq:Reff_Leff} describe the origin of the observed flux dependence of \(Q\) and \(f_{\mathrm r}\). Our goal is to infer \(R_{\mathrm{SQ}}\) and \(L_{\mathrm{SQ}}\) from the measured variations
\begin{align*}
\delta Q &= Q - Q_0,\\
\delta f_{\mathrm r} &= f_{\mathrm r} - f_{\mathrm r,0},
\end{align*}
where \(Q_0\) and \(f_{\mathrm r,0}\) denote, respectively, the quality factor and resonance frequency obtained with the SQUID unbiased; their values are listed in Table~\ref{tab:resonator_params}. Treating \(R_{\mathrm{SQ}}\) and \(L_{\mathrm{SQ}}\) as small parameters, we expand Eqs.~\eqref{eq:Reff_Leff} about \(R_{\mathrm{SQ}}=L_{\mathrm{SQ}}=0\). At this point,
\begin{align*}
R\big|_0 &= R_0,\\
L\big|_0 &= L_0 - \frac{M_{\mathrm r}^2}{L_{\mathrm r}}.
\end{align*}
To leading order in \(L_{\mathrm{SQ}}\),
\begin{equation*}
\delta L = \frac{M_{\mathrm r}^2}{L_{\mathrm r}^2}L_{\mathrm{SQ}} .
\end{equation*}
Since \(f_{\mathrm r}=1/(2\pi\sqrt{LC_0})\), the corresponding fractional frequency shift is
\begin{equation}
\label{eq:f_L_var}
\frac{\delta f_{\mathrm r}}{f_{\mathrm r,0}} = -\frac{1}{2}\frac{\delta L}{L\big|_0} = -\frac{M_{\mathrm r}^2}{2L_{\mathrm r}^2 L\big|_0}L_{\mathrm{SQ}}.
\end{equation}
Thus,
\begin{equation}
\label{eq:Lsq}
L_{\mathrm{SQ}} = -2\frac{L_{\mathrm r}^2}{M_{\mathrm r}^2}\left(L_0-\frac{M_{\mathrm r}^2}{L_{\mathrm r}}\right)\frac{\delta f_{\mathrm r}}{f_{\mathrm r,0}} .
\end{equation}
Similarly, \(Q\) is given by
\begin{equation*}
Q = \frac{1}{R}
\sqrt{\frac{L}{C_0}} .
\end{equation*}
Therefore,
\begin{equation*}
\frac{\delta Q}{Q_0} = -\frac{\delta R}{R\big|_0} + \frac{1}{2} \frac{\delta L}{L\big|_0} = -\frac{\delta R}{R\big|_0} - \frac{\delta f_{\mathrm r}}{f_{\mathrm r,0}},
\end{equation*}
where Eq.~\eqref{eq:f_L_var} was used. Using
\begin{equation*}
\delta R = \frac{M_{\mathrm r}^2}{L_{\mathrm r}^2}R_{\mathrm{SQ}},
\end{equation*}
we obtain
\begin{equation*}
R_{\mathrm{SQ}} = -\frac{L_{\mathrm r}^2}{M_{\mathrm r}^2} R_0 \left(\frac{\delta Q}{Q_0} + \frac{\delta f_{\mathrm r}}{f_{\mathrm r,0}}\right).
\end{equation*}
Since \(\delta f_{\mathrm r}/f_{\mathrm r,0}\ll \delta Q/Q_0\) for these measurements, we approximate this expression as
\begin{equation}
\label{eq:Rsq}
R_{\mathrm{SQ}} \simeq -\frac{L_{\mathrm r}^2}{M_{\mathrm r}^2} R_0 \frac{\delta Q}{Q_0}.
\end{equation}
For example, at the largest observed \(Q\) deviation, \(\delta Q/Q_0\simeq 0.3\), whereas \(\delta f_{\mathrm r}/f_{\mathrm r,0}\simeq 2\times 10^{-7}\). The approximation \(R_{\mathrm{SQ}}\ll \omega(L_{\mathrm r}+L_{\mathrm{SQ}})\) used in deriving Eq.~\eqref{eq:Z} is verified a posteriori: the largest inferred value, \(R_{\mathrm{SQ}}\sim 8~\mathrm{m}\Omega\), shown in Fig.~\ref{fig:Lsq_and_Rsq_vs_flux}, is about \(8\times10^{-3}\) of \(\omega(L_{\mathrm r}+L_{\mathrm{SQ}})\), so the correction is negligible.

Equations~\eqref{eq:Lsq} and~\eqref{eq:Rsq} relate the measured shifts in resonance frequency and \(Q\), shown in Fig.~\ref{fig:Q_and_fr_dVdPhi_vs_flux}, to the flux-dependent SQUID self-inductance and resistance, shown in Fig.~\ref{fig:Lsq_and_Rsq_vs_flux}. For \(L_{\mathrm{SQ}}\), however, there is an additional caveat: the resonance-frequency shift is small and is obtained by comparing measurements with the SQUID unbiased and biased. These two measurements use different output settings, which can introduce a small parasitic impedance change and therefore an unknown overall offset in \(\delta f_{\mathrm r}\). Consequently, the inferred \(L_{\mathrm{SQ}}\) is determined only up to an additive constant. For this reason, in Fig.~\ref{fig:Lsq_and_Rsq_vs_flux} we show \(L_{\mathrm{SQ}}-\langle L_{\mathrm{SQ}}\rangle_{\Phi_0}\), where
\begin{equation}
\left\langle L_{\mathrm{SQ}}\right\rangle_{\Phi_0} = \frac{1}{\Phi_0} \int_{0}^{\Phi_0} L_{\mathrm{SQ}}(\Phi)\,d\Phi .
\end{equation}
Previous SQUID measurements indicate that the SQUID dynamic inductive response changes sign over one flux period \cite{Hilbert1985}. Using the measured extrema of \(L_{\mathrm{SQ}}-\left\langle L_{\mathrm{SQ}}\right\rangle_{\Phi_0}\), which range from approximately \(-16~\mathrm{nH}\) to \(33~\mathrm{nH}\), we estimate
\begin{equation}
\label{eq:Lsq_boundaries}
-33~\mathrm{nH} \lesssim \left\langle L_{\mathrm{SQ}}\right\rangle_{\Phi_0} \lesssim 16~\mathrm{nH}.
\end{equation}
\bibliographystyle{apsrev4-2}
\bibliography{refs}
\end{document}